\documentclass[12pt]{article}
\usepackage{amssymb}
\usepackage{graphicx,color}
\usepackage{amsmath}
\usepackage{enumerate}
\usepackage{epsfig}
\usepackage[english]{babel}
\usepackage{latexsym}

\def\sod{{\leavevmode\setbox1=\hbox{d}%
\hbox to 1.05\wd1{d\kern-0.4ex{\char039}\hss}}}
\def\sot{{\leavevmode\setbox1=\hbox{t}%
\hbox to \wd1{t\kern-0.5ex{\char039}\hss}}}
\def\sol{l\kern-0.3ex\raise0.1ex\hbox{'}\kern-0.10ex}
\def\soL{L\kern-0.8ex\raise0.1ex\hbox{'}\kern0.1ex}
\begin{document}
\title{Effect of Compressibility on the Annihilation Process}

\author{M. Hnatich$^{1,2}$, J. Honkonen$^{3}$, T. Lu\v{c}ivjansk\'y$^{1,2}$}

\maketitle\mbox{ }
\\
$^{1}$ Institute of Experimental
Physics, Slovak Academy of Sciences, Watsonova 47, 040 01
Ko\v{s}ice, Slovakia,\\
$^{2}$ Faculty of Sciences, P.J. \v{S}afarik
University, Moyzesova 16, 040 01 Ko\v{s}ice, Slovakia,\\
$^3$
Department of Military Technology, National Defence University,
P.O.~Box~7, 00861, Helsinki, Finland.\\
\begin{abstract}
Annihilation processes, where the reacting particles are influenced by some
external advective field, are one of the simplest examples of nonlinear
 statistical systems. This type of processes can be observed in miscellaneous chemical,
 biological or physical systems. In low space dimensions usual description by means of kinetic
  rate equation is not sufficient and the
 effect of density fluctuations must be taken into account. Using perturbative
 renormalization group we study the influence of random velocity field on the kinetics
  of single-species annihilation reaction at and below its critical dimension $d_c=2$.
  The advecting velocity field is modelled by the self-similar in space Gaussian
   variable finite correlated in time (Antonov-Kraichnan model).
  Effect of the compressibility of velocity field is taken into account and the
   model is analyzed near its critical dimension by means of three-parameter
    expansion in $\epsilon,\Delta$ and $\eta$. Here $\epsilon$ is the deviation
    from the Kolmogorov scaling, $\Delta$ is the deviation from the (critical)
    space dimension $2$ and $\eta$ is the deviation from the parabolic dispersion
     law. Depending on the value of these exponents and the value of compressiblity
      parameter $\alpha$, the studied model can exhibit various asymptotic
      (long-time) regimes corresponding to the infrared (IR) fixed points of the renormalization
       group. The possible regimes are summarized and the decay rates for the mean particle
        number are calculated in the leading order of the perturbation theory.
\end{abstract}
{\section{Introduction} \label{eq:anni3_intro}}
Variety of chemical reactions occur in fluid environment and
represents very important subject in various chemical, biological or
physical systems \cite{Der95,Kro93,Tel05}. In these processes the reacting particles are affected not
only by the diffusion motion but also by the external fluid flow.
The usual approach to such reactive flows is based
 on some combination of reaction-transport equations.

 In this work we will concentrate on the study of
 the annihilation reaction $\textit{A} +\textit{A} \rightarrow\varnothing$, which
 is the paradigmatic model for the reaction-diffusion processes.
 For this type of reaction in low space dimensions the usual
description by means of kinetic rate equation is not sufficient and
the effect of density fluctuations must be taken into the account \cite{Lee}.
It can be shown that the upper critical dimension for this process, even in the
 absence of advective flow, is two  due to the density fluctuations. A renormalization group treatment
  was successfully applied for different choices of reactive flow, e.g.
 time-independent random drift \cite{Deem}, velocity field described by
 stochastic Navier-Stokes equation \cite{Hnatic} or short-range correlated potential disorder \cite{Richard}.
 Often large differences are observed in relation to the experiment
 and their origin can be caused by the presence of large-scale anisotropies, effect of compressibility
or parity violation. There are important differences \cite{Celani99,Benzi12} between advection of scalar
quantities like density on one hand and like tracer (temperature, concentration)
 on the other by compressible versus
incompressible flow. It was shown that compressibility could lead to the slowing of
transport process for scalar admixture and also to the enhancement of intermittent phenomena.
These effects might be understood
as a result of inhibition of separation between particle trajectories and therefore we expect
that reacting particles would spent effectively more time in the mutual vicinity than
in the incompressible case. Hence in this case we expect the faster decay rate
than for the incompressible case. Hence it would be desirable to support such naive
picture in more quantitative manner. In order to do that we
  will apply Antonov-Kraichnan model \cite{Kraichnan68,Antonov} for describing advection of reactive $A$
   particles and we will present renormalization group (RG) study in the vicinity of
 its critical dimension $d_c=2$. In the one-loop order all relevant physical quantities are
calculated and contrary to the incompressible case it is found out, that already in the one-loop approximation
fluctuations of the velocity field affect the renormalization of the rate constant.

 After brief description of the model in Sec. \ref{sec:anni3_model},
  results of RG calculations are presented in Sec. \ref{sec:anni3_rgfunc}. In
 Sec. \ref{sec:anni3_stable_regimes} possible large-scale regimes are listed and their physical interpretation
is concluded in Sec. \ref{sec:anni3_conclusion}.
{\section{Field-theoretic model} \label{sec:anni3_model}}
Field-theoretic action for the annihilation reaction process
$A+A \xrightarrow{K_0} \varnothing$ can be
obtained in the straightforward way \cite{Tau05}
 employing Doi approach \cite{Doi}.
  It can be written  in the following standard form
\begin{eqnarray}
  S_{1} & = &  - \int^{\infty}_{0} {\mathrm d}t \int {\mathrm d} {\bf x}
   \biggl\{ \psi^\dagger \partial_{t} \psi
   -D_{0}\psi^\dagger\nabla^{2}\psi
  +    \lambda_{0} D_{0}[2\psi^\dagger+(\psi^\dagger)^{2}]\psi^{2} \biggl\} +\nonumber \\
  & + &   n_{0}\int {\mathrm d}{\bf x}\mbox{ } \psi^\dagger({\bf x},0) \, ,
  \label{eq:S_1}
\end{eqnarray}
 where $D_0$ is the diffusion constant of reacting particles, and because of dimensional reasons
 we have rewritten product $\lambda_0 D_0$ instead of the rate constant $K_0$. Last term in the action
 stands for the initial conditions, which are traditionally chosen in the form
 of Poisson distribution.

In compressible velocity field, there are two types of diffusion-advection problems: advection of a density field
 and advection of a tracer field \cite{Landau}. Here, the case of advection of a density field will be analyzed \cite{Antonov01}.
 In the action (\ref{eq:S_1}) this corresponds to the replacement
 $\psi^\dagger \partial_{t} \psi\to \psi^\dagger\left[\partial_{t}\psi +(\nabla.{\bf v}\psi)\right]$,
where ${\bf v}={\bf v}({\bf x},t)$ is the
 advecting velocity field. According to \cite{Antonov}  let us assume that ${\bf v}$ is a random Gaussian variable
 with zero mean and the correlator (in the frequency-momentum representation)
\begin{equation}
  \langle v_{0i} v_{0j}\rangle_0 = \frac{g_0 D_0^3 k^{2-2\Delta-2\epsilon-2\eta}}{\omega^2+(u_0D_0 k^{2-\eta})^2}
  [P_{ij}({\mathbf k})+\alpha Q_{ij}({\mathbf k})],
  \label{correlator_v}
\end{equation}
where $g_0$ is the coupling constant, the exponents $\epsilon$,$\Delta$ and
$\eta$ play the role of small expansion parameters. They could be
regarded as an analog of the expansion parameter $\epsilon=4-d$ used
in the theory of critical phenomena. However, in this
paper $\epsilon$ should be understood as deviation of exponent of
the power law from that of the Kolmogorov scaling \cite{Frisch},
whereas $\Delta$ is defined as the deviation from the space
dimension two via relation $d=2+2\Delta$, and the exponent $\eta$ is
related to the reciprocal of the correlation time at the wave number
$k$. The parameter $u_0$ serves for labeling of the fixed
points and it can be interpreted as a ratio of velocity correlation time
and the scalar turnover time \cite{Adzhemyan2}. 
In (\ref{correlator_v}) besides the standard (incompressible)  transverse
projection operator $P_{ij}({\bf k}) = \delta_{ij}-k_ik_j/k^2$ the
 longitudinal projector $Q_{ij}({\mathbf k}) = k_i k_j/k^2$ has been introduced.
The positive (necessary for positive definiteness of the correlator
 $\langle v v\rangle$) parameter $\alpha$ represents the degree of compressibility.
The incompressible case is obtained by the setting $\alpha=0$. \\
 The Antonov-Kraichnan model for the advection field ${\bf v}$ contains two cases of special interest:
\begin{enumerate}[(a)]
 \item in the limit $u_0 \rightarrow \infty, g_0' \equiv g_0/u_0^2 = const$ we get the 'the rapid-change model'
  $ D_v(\omega,{\bf k}) \rightarrow g_0' D_0 k^{-2-2\Delta-2\epsilon+\eta}$,
  which is characterized by the white-in-time nature of the velocity correlator.
 \item limit $u_0 \rightarrow 0, g_0''\equiv g_0/u_0 = const$ corresponds to the case of a frozen velocity field
  $ D_{v}(\omega,{\bf k}) \rightarrow g_0'' D_0^2 \pi \delta(\omega) k^{2\Delta-2\epsilon}$,
 when the velocity field is quenched (time-independent).
\end{enumerate}
The averaging procedure with respect to the velocity field ${\bf v}(x)$ 
may be performed with the aid of the following action functional
\begin{equation}
  S_2 = -\frac{1}{2}\int {\mathrm d}t {\mathrm d}{\bf x}
  \int {\mathrm d}t'{\mathrm d}{\bf x'} \mbox{ }
  {\bf v}(t,{\bf x}) D_v^{-1}(t-t',{\bf x}-{\bf x'}) {\bf v}(t',{\bf x'}),
  \label{eq:S_2}
\end{equation}
where $D_v^{-1}$ is the inverse correlator (\ref{correlator_v}) (in
the sense of the Fourier transform). The expectation value of any
relevant physical observable may be calculated using the complete
weight functional $\mathcal{W}(\psi^\dagger,\psi,{\bf v}) = {\rm
e}^{S_1+S_2}$, where $S_1$ and $S_2$ are the action functionals
(\ref{eq:S_1}) and (\ref{eq:S_2}).

{\section{UV renormalization} \label{sec:anni3_rgfunc}}
The inclusion of longitudinal part $Q$ into the correlator for
velocity field  does not affect the renormalization group
analysis developed for the such model \cite{Hnatic11}. Therefore we just mention main
 steps of theoretical description and deviations from it caused by compressibility violation.
All canonical dimensions of fields and parameters
are listed in Tab. \ref{tab:anni2_canon_dim}.
\begin{table}[h!]
\centering
\begin{tabular}{|c|c|c|c|c|c|c|c|c|}
  \hline
  $Q$ & $\psi$ & $\psi^\dagger$ & $v$ & $D_0$ & $u_0$ & $\lambda_0$ & $g_0$ & $\alpha,g,u,\lambda$ \\ \hline
  $d_Q^k$ & $d$ & $0$ & $-1$ & $-2$ & $\eta$ & $-2\Delta$ & $2\epsilon+\eta$ & 0 \\ \hline
  $d_Q^\omega $& $0 $ & $0$ & $1$ & $1$ & $0$ & $0$ & $0$ & 0 \\ \hline
  $d_Q $& $d$ & $0$ & $1$ & $0$ & $\eta$ & $-2\Delta$ & $2\epsilon+\eta$ & 0 \\ \hline
\end{tabular}
  \caption{Canonical dimensions for the parameters and fields of the model}
  \label{tab:anni2_canon_dim}
\end{table}
The only difference with the incompressible case is that now the velocity
field has to be renormalized \cite{Antonov}.
Following \cite{Hnatic,Antonov,Hnatic11} it is easy to prove that the model under consideration is
multiplicatively renormalizable and can be made UV finite by the following renormalization
 prescription
\begin{equation}
   D_0 = D Z_D,\mbox{ } g_0=g\mu^{2\epsilon+\eta}Z_g,\mbox{ } u_0=u\mu^\eta Z_u,\mbox{ }
   \lambda_0=\lambda\mu^{-2\Delta}Z_D^{-1}Z_\lambda,\mbox{ } {\mathbf v}_0 = {\mathbf v }Z_v
   \label{eq:anni3_ren_rel1}
\end{equation}
with the additional constraints between them
\begin{equation}
  Z_gZ_D^3=1, \quad  Z_uZ_D=1, \quad Z_g Z_D^3 = Z_{ v}^2,
     \label{eq:anni3_ren_rel2}
\end{equation}
which are the consequences of the absence of renormalization of non-local term (\ref{correlator_v}).
 Non-local character of the velocity correlator is caused by the nontrivial correlations in momentum and frequency scales.
The total renormalized action can be written as
\begin{eqnarray}
   S_R(\psi^\dagger,\psi,{\mathbf v}) & = & -
   \int^{\infty}_{0} {\mathrm d}t \int {\mathrm d}{\bf x}
   \biggl\{ \psi^\dagger\partial_t\psi -\psi^\dagger DZ_D \nabla^2\psi +\psi^\dagger Z_v
   ({\bf\nabla}.{\mathbf v}\psi)]
    - \nonumber\\
    & - & Z_\lambda D\lambda [2\psi^\dagger+\psi^{\dagger 2}]\psi^2 \biggl\}
    -\int {\mathrm d}t {\mathrm d}{\bf x} \int {\mathrm d}t'{\mathrm d}{\bf x'} \mbox{ }
     \frac{v D_v^{-1}v}{2} + \nonumber \\
     & + & n_{0}\int {\mathrm d}{\bf x}\mbox{ } \psi^\dagger({\bf x},0).
   \label{eq:anni3_renorm_action}\nonumber
\end{eqnarray}
The perturbative calculation of the renormalization constants in dimensional regularization with the
use minimal subtraction (MS) scheme is straightforward \cite{Zinn}.
We restrict ourselves to the first order in perturbation theory
and this approximation already contains
first nontrivial effect of the compressibility. It
 can be seen from the perturbation expansion of the one-particle irreducible
  function $\Gamma_{\psi^{\dagger}\psi^2}$ (known as interaction vertex)
\begin{equation}
  \Gamma_{\psi^\dagger\psi^2}|_{\omega=0,p^2=0} =  -4 D\lambda Z_\lambda\mu^{-2\Delta} +\frac{1}{2}
  \raisebox{-4.3ex}{ \epsfysize=1.7truecm
  \epsffile{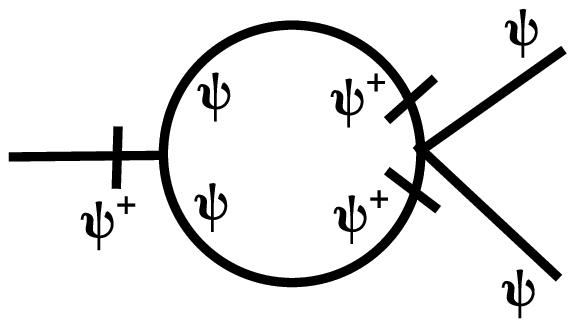}} + \frac{1}{2}
  \raisebox{-4.75ex}{ \epsfysize=1.9truecm
  \epsffile{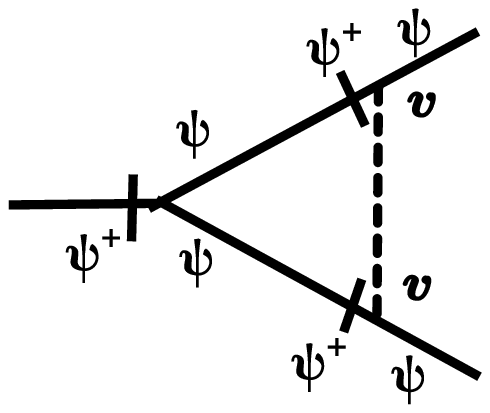}}\,\,,
  \label{eq:anni3_trilinear}
\end{equation}
where higher order terms in coupling constants were neglected.  The first graph on the
  r.h.s. of (\ref{eq:anni3_trilinear}) physically represents
process of density fluctuations that account for annihilating of two particles.
It is easy to see that transverse character (equivalent statement to the
incompressibility condition for velocity field ${\mathbf v}$) of
propagator $\langle v v\rangle$ leads to UV divergent contribution
of the second graph in (\ref{eq:anni3_trilinear}), whereas for the models
with incompressible velocity field ${\mathbf v}$ it leads to UV convergent contribution.
Physically this graph can be interpreted as an attracting process
that brings together particles into a sink of compression, which can lead to
effective increase of the reaction rate as will be pointed later.
In the MS scheme the renormalization constants in one-loop calculation obtain the
following form
\begin{eqnarray}
  Z_D & = & 1-\frac{g}{16\pi u(1+u)\epsilon}\biggl[1+\alpha-\frac{2\alpha}{1+u}\biggl], \\
  Z_v & = & 1 + \frac{\alpha g}{16\pi u(1+u)^2\epsilon},\\
  Z_\lambda & = & 1 - \frac{\lambda}{4\pi\Delta} - \frac{\alpha g}{16\pi u(1+u)\epsilon},
  \label{eq:anni3_renorm_constants}
\end{eqnarray}
and from relations (\ref{eq:anni3_ren_rel2}) constants $Z_u$ and $Z_g$ can be
calculated.
Limiting case $\alpha=0$ agrees with the results  (to the one-loop precision)
 for incompressible case \cite{Hnatic11} and case $g=0$ leads to the presence
 of only density fluctuations \cite{Lee}.

From the relations (\ref{eq:anni3_ren_rel1}) and (\ref{eq:anni3_ren_rel2}) the  beta functions
$\beta_g,\beta_u$ and $\beta_\lambda$  are obtained via the standard
definition $\beta_g = \mu \partial_\mu g |_0$ (the subscript ''0'' refers to partial derivatives taken at
fixed values of the bare (unrenormalized) parameters)
\begin{equation}
  \beta_g = g[-2\epsilon-\eta+3\gamma_D-2\gamma_{\mathbf v}],\quad
  \beta_u = u[-\eta+\gamma_D],\quad
   \beta_\lambda = \lambda[2\Delta-\gamma_\lambda+\gamma_D] ,
   \label{eq:anni3_beta_func}
\end{equation}
where the anomalous dimensions $\gamma_F$ are defined as \cite{Zinn}
\begin{equation}
  \gamma_F = \mu\partial_\mu \ln Z_F |_0 =(\beta_g\partial_g +\beta_u\partial_u+\beta_\lambda\partial_\lambda)
  \ln Z_F .
  \label{eq:anni3_anomalous}
\end{equation}
It was conjectured \cite{Antonov,Adzhemyan2,Ant99}, that up to the two-loop
calculations there is no direct influence of the parameter $\eta$ on the
 anomalous dimensions. Hence one can use in the actual calculations of the Feynman graphs
 different values of  $\eta$. In our calculations the simplest choice $\eta=0$
  was applied. Finally substituting (\ref{eq:anni3_renorm_constants}) into
  the definition (\ref{eq:anni3_anomalous})  the anomalous dimensions are obtained
\begin{eqnarray}
  & & \gamma_D = \frac{g}{8\pi u(1+u)}\biggl[1+\alpha-\frac{2\alpha}{u+1}\biggl],\quad
  \gamma_{ v} = -\frac{\alpha g}{8\pi u(1+u)^2},\nonumber\\
  & & \gamma_\lambda = \frac{\alpha g}{8\pi u(1+u)}-\frac{\lambda}{2\pi}.
  \label{eq:anni3_anomal_dim}\nonumber
\end{eqnarray}
{\section{IR stable regimes} \label{sec:anni3_stable_regimes}}
We are interested in the IR asymptotics of small momentum ${\bf p}$ and frequencies $\omega$
of the renormalized functions or, equivalently, large relative distances and time differences in
the $(t,{\bf x})$ representation. Such a behavior is governed by the IR-stable fixed point
$g^*=(g_1^*,u^*,\lambda^*)$, which are determined as zeroes of the $\beta$ functions (\ref{eq:anni3_beta_func}):
$\beta(g^*)=0$  . It is said, that the fixed point $g^*$ is IR stable, if real parts of all eigenvalues
of the matrix $\Omega_{ij}\equiv \partial \beta_i/\partial g_j|_{g=g^*}; i,j\in\{g,u,\lambda\}$
are strictly positive.

 The simplest way to find the average number density $n(t)=\langle\psi(t)\rangle$  is to calculate it from
the stationarity condition of the functional Legendre transform \cite{Vasiliev}
 of the generating functional obtained by replacing the unrenormalized
  action by the renormalized one in the weight functional \cite{Hnatic}. This is a convenient way to
avoid any summing procedures used \cite{Lee} to take into account the higher-order terms in $n_0$.
For a spatially homogenous solution this leads to the rate equation
with the initial condition $n(0)=n_0$ for the average number
density $n(t)=\langle \psi(t)\rangle$
\begin{equation}
  \label{anni1_NLsolution}
  n(t)=\frac{n_0}{1+2{\lambda} u D t\mu^{-2\Delta}n_0}\,,\nonumber
\end{equation}
where $n_0$ is the initial number density.
Since the fields $\psi$ and $\psi^\dagger$ are not renormalized,
the Callan-Symanzik equation for the mean particle number is easily obtained
by the standard procedure \cite{Hnatic,turbo}
\begin{equation}
  \biggl[(2-\gamma_D)t\frac{\partial}{\partial t}+\sum_{g} \beta_g \frac{\partial}{\partial g}-
  dn_0 \frac{\partial}{\partial n_0}+d \biggr] n\left(t,\mu,D,n_0,g\right)=0
  \label{eq:anni1_Callan}\nonumber
\end{equation}
Solving it by the means of characteristics it can be shown (details will be
published elsewhere \cite{HnaticEur}) that the value of
the decay exponent $\beta$ defined through the asymptotic
 relation : $n(t) \underset{t\rightarrow\infty}{\sim} t^{-\beta} $ is given by the expression
\begin{equation}
  \label{eq:anni1_defalpha}
  \beta= 1+\frac{\gamma_\lambda^*}{\displaystyle 2-\gamma_D^*}.\nonumber
\end{equation}
 Note that in contrast to the previous studies \cite{Hnatic,Hnatic11} here we have to deal only
 with the three-charges' $\{g,u,\lambda\}$ theory.
 Passive character (no backward influence on the advecting field) of the reacting particle
  manifests itself also in the relations
  $\partial_\lambda \beta_g= \partial_\lambda \beta_u=0$ resulting from
  (\ref{eq:anni3_renorm_constants}) and  (\ref{eq:anni3_beta_func}). These relations greatly simplify
 calculation of the eigenvalues of $\omega_{ij}$ matrix.

Detailed analysis of fixed point structure reveals that studied system can exhibit one
of ten possible IR regimes listed below.
First let us consider the ''rapid-change model'' ($u\rightarrow\infty$) .
Introducing convenient variables $w=1/u, g'=g/u^2$, the corresponding $\beta$ functions
can be written in the form
\begin{equation}
  \beta_{g'} = g'[-2\epsilon+\eta + \gamma_D -2\gamma_v],\quad
  \beta_w = w[\eta-\gamma_D],\quad
  \beta_\lambda = \lambda[2\Delta-\gamma_\lambda+\gamma_D],
  \label{eq:anni3_beta_func_rapid}\nonumber
\end{equation}
where anomalous dimensions are
\begin{equation}
  \gamma_D  =  \frac{g'}{8\pi (1+w)} \biggl[1+\alpha-\frac{2\alpha w}{1+w} \biggl],\mbox{ }
  \gamma_v  =  -\frac{\alpha g'}{8\pi (1+w)^2},\mbox{ }
  \gamma_\lambda  =  \frac{\alpha g'}{8\pi (1+w)} - \frac{\lambda}{2\pi}.
  \label{eq:anni3_anomal_dim_rapid}\nonumber
\end{equation}
The ''rapid-change model'' corresponds to the fixed point with the value $w^*=0$.
 In this case four stable IR fixed points can be realized:
\begin{eqnarray*}
    \mbox{\bf FP 1: } & & {g'}^*=0,\quad \lambda^* = 0; \nonumber \\
    & & \Omega_1 = \eta-2\epsilon,\quad\Omega_2=\eta,\quad \Omega_3 = 2\Delta; \nonumber \\
    & & \beta = 1;\nonumber\\
    \mbox{\bf FP 2: } & & {g'}^*=0,\quad\lambda^* = -4\pi\Delta; \nonumber \\
    & & \Omega_1 = \eta-2\epsilon,\quad \Omega_2=\eta,\quad \Omega_3=-2\Delta; \nonumber \\
    & & \beta=1+\Delta;\nonumber\\
    \mbox{\bf FP 3: } & &  {g'}^* = \frac{8\pi(2\epsilon-\eta)}{1+\alpha},\quad \lambda^* = 0; \nonumber \\
    & & \Omega_1=2\epsilon-\eta
    ,\quad \Omega_2=2\eta-2\epsilon,\quad \Omega_3=2\Delta+\frac{2\epsilon-\eta}{1+\alpha};     \nonumber \\
    & & \beta = \frac{2\alpha+2-2\epsilon+\eta}{(1+\alpha)(2-2\epsilon+\eta)};
    \nonumber\\
   \mbox{\bf FP 4: } & &  {g'}^* = \frac{8\pi(2\epsilon-\eta)}{1+\alpha}, \quad
                \lambda^* = 2\pi\biggl(  2\Delta+\frac{-2\epsilon+\eta}{1+\alpha} \biggl); \nonumber \\
   & & \Omega_1 = 2\epsilon-\eta,\quad \Omega_2=2\eta-2\epsilon,\quad
    \Omega_3=-2\Delta-\frac{2\epsilon-\eta}{1+\alpha}; \nonumber\\
   & & \beta = \frac{2+2\Delta}{2-2\epsilon+\eta} .
\end{eqnarray*}
For the analysis of the regime $u\rightarrow 0$ (quenched velocity field)
 we introduce the new variable $g''\equiv g/u$. Hence the corresponding $\beta$ functions have the form
\begin{equation}
  \beta_{g''} = g''[-2\epsilon + 2\gamma_D -2\gamma_v]\,,\quad
  \beta_u = u[-\eta+\gamma_D]\,,\quad
  \beta_\lambda = \lambda[2\Delta-\gamma_\lambda+\gamma_D]\,.
  \label{eq:anni3_beta_func_frozen}\nonumber
\end{equation}
and anomalous dimensions are given as
\begin{equation}
  \gamma_D  =  \frac{g''}{8\pi (1+u)} \biggl[1+\alpha-\frac{2\alpha}{1+u} \biggl],\quad
  \gamma_v  =  -\frac{\alpha g''}{8\pi (1+u)^2},\quad
  \gamma_\lambda =  \frac{\alpha g''}{8\pi (1+u)} - \frac{\lambda}{2\pi}.
  \label{eq:anni3_anomal_dim_frozen}\nonumber
\end{equation}
The quenched regime corresponds to the $u^*=0$
 and also in this case there are four possible IR stable fixed points:
\begin{eqnarray*}
   \mbox{\bf FP 5: } & & {g''}^*= 0,\quad \lambda^* = 0; \nonumber \\
   & & \Omega_1=-2\epsilon,\quad \Omega_2=-\eta,\quad \Omega_3=2\Delta \\
   & & \beta=1;\nonumber\\
   \mbox{\bf FP 6: } & & {g''}^*= 0,\quad \lambda^*=-4\pi\Delta;   \nonumber\\
   & & \Omega_1=-2\epsilon,\quad \Omega_2=-\eta,\quad \Omega_3=-2\Delta; \nonumber \\
      & & \beta=1+\Delta ;\nonumber\\
   \mbox{\bf FP 7: } & & {g''}^*=8\pi\epsilon,\quad \lambda^* = 0;  \nonumber\\
   & & \Omega_1=2\epsilon,\quad \Omega_2=-\eta+\epsilon(1-\alpha),\quad
    \Omega_3 =2\Delta+\epsilon(1-2\alpha); \nonumber \\
       & & \beta=\frac{2-(1-2\alpha)\epsilon}{2-(1-\alpha)\epsilon} ;\nonumber\\
   \mbox{\bf FP 8: } & &  {g''}^*=8\pi\epsilon, \quad \lambda^* =  2\pi[-2\Delta+\epsilon(2\alpha-1)]; \nonumber\\
   & & \Omega_1=2\epsilon,\quad \Omega_2=-\eta+(1-\alpha)\epsilon,\quad
  \Omega_3 =-2\Delta+(2\alpha-1)\epsilon; \nonumber\\
     & & \beta=\frac{2+2\Delta}{2-(1-\alpha)\epsilon}.\nonumber
\end{eqnarray*}
The nontrivial case occurs when no special
choice for parameter $u$ is considered, i.e. let's consider $u$ finite and non-zero.
Solving equations (\ref{eq:anni3_beta_func}) for $u\neq 0,g\neq 0$ the
following values for the coordinates of the fixed point are obtained
\begin{equation}
  \frac{g^*}{8\pi u^*(1+u^*)} = \frac{2\epsilon-\eta}{1+\alpha},\quad u^*=-1+\frac{\alpha(\eta-2\epsilon)}{(1+\alpha)(\eta-\epsilon)}\,\,.
  \label{eq:anni3_nontrivial}
\end{equation}
The two possible regimes are distinguished by the value
of the coordinate $\lambda^*$. Fixed point with zero value is given by
\begin{equation}
  \mbox{\bf FP 9: } \lambda^* = 0, \quad
  \beta =\frac{2-\eta+\alpha(1+\epsilon-\eta)}{(1+\alpha)(2-\eta)} \nonumber
\end{equation}
and is stable in the region
\begin{equation}
 (1-\alpha)\epsilon < \eta < \epsilon, \quad
 \Delta+\frac{(1+2\alpha)\eta}{2(1+\alpha)} > \frac{\alpha\epsilon}{1+\alpha}\, .
 \label{fp9_ros}
\end{equation}
The fixed point with non-zero value of $\lambda^*$ is given by
\begin{equation}
  \mbox{\bf FP 10: }
   \lambda^*=-4\pi\Delta + \frac{2\pi\alpha\epsilon}{1+\alpha}-2\pi\eta\frac{1+2\alpha}{1+\alpha},\quad
   \beta =\frac{2+2\Delta}{2-\eta}\,,\nonumber
\end{equation}
and is stable in the region
\begin{equation}
 (1-\alpha)\epsilon < \eta < \epsilon, \quad
  \Delta+\frac{(1+2\alpha)\eta}{2(1+\alpha)} < \frac{\alpha\epsilon}{1+\alpha} \,.
  \label{fp10_ros}
\end{equation}
{\section{Conclusions} \label{sec:anni3_conclusion}}
Fixed points ${\bf 1}$ and ${\bf 5}$ corresponds
to the non-interacting (mean-field or Gaussian) theory and thus their predictions should agree
 with the ones of rate equation approach, which is indeed the case \cite{Lee}.

From the fixed points' structure some physical consequences can be deduced. First we see, that
compressibility has direct influence on the value of decay exponent (see {\bf FP 3}, {\bf FP 7-9}).
For some regimes  ({\bf FP 3}, {\bf FP 7}) it could lead to the enhancement (for both of them $\beta>1$) of the reaction process
  compared to the corresponding regimes
for incompressible case \cite{Hnatic,Hnatic11}. As was already pointed this fact can be explained by
the presence of compressible sinks into which particles are attracted (see also Sec. 4 in \cite{Bouchaud}). However we also
 observe that when both density fluctuations and compressibility are relevant ({\bf FP 4},  {\bf FP 10}), the density tends
to suppress the influence of compressibility.

The ``real problem'' corresponds to the choice $\epsilon=\eta=4/3$, which leads to the famous Kolmogorov ``five-thirds law''
 \cite{Frisch} for the spatial velocity statistics. It is easy to see, that this regime can be realized
 either by the fixed point {\bf FP 9} or {\bf FP 10} depending on the value of parameter $\Delta$, or
 equivalently on the space dimension $d$.
 From (\ref{fp9_ros}) and (\ref{fp10_ros}) we see
  that there is a "critical" value $\Delta_c=-2/(3+3\alpha)$ for
 the parameter $\Delta$, above which {\bf FP 9} is stable, whereas below it {\bf FP 10} is stable.
 Because $\alpha$ is positive quantity, a parameter $\Delta_c$ is negative. Therefore in the vicinity of
 the space dimension two ($\Delta=0$) regime represented by the fixed point {\bf FP 9} should be realized with
the decay exponent $\beta=(1+3\alpha)/(1+\alpha)$.
 Thus we can conclude, that compressibility has a profound effect on the large-scale asymptotic behavior of the
  annihilation process and leads to the enhancement of it near its critical dimension $d_c=2$. On the other hand
  density fluctuations leads again to the suppression of compressibility (as can be
  seen by direct numerical inspection of exponent $\beta$ for {\bf FP 10}). \\

The work was supported by VEGA grant 1/0222/13 of the Ministry 
of Education, Science, Research and Sport of the Slovak Republic, by
Centre of Excellency for Nanofluid of IEP SAS. This
article was also created by implementation of the Cooperative phenomena
and phase transitions in nanosystems with perspective utilization in nano-
and biotechnology project No 26220120033 and No 26110230061. Funding
for the operational research and development program was provided by the
European Regional Development Fund.

\end{document}